\documentclass[10pt, conference]{IEEEtran}
\IEEEoverridecommandlockouts
\usepackage{bm}
\usepackage{braket}
\usepackage{cite}
\usepackage{amsmath,amssymb,amsfonts}
\usepackage{algorithmic}
\usepackage{graphicx}
\usepackage{textcomp}
\usepackage{xcolor}
\usepackage{booktabs}
\usepackage{url}
\usepackage{float}
\usepackage{stfloats}

\def\BibTeX{{\rm B\kern-.05em{\sc i\kern-.025em b}\kern-.08em
    T\kern-.1667em\lower.7ex\hbox{E}\kern-.125emX}}
\begin{document}

\title{A Novel Parallel QCNN Architecture with Efficient Classical Simulability}

\author{\IEEEauthorblockN{Lawrence Nguyen}
\IEEEauthorblockA{\textit{Department of Electrical Engineering} \\
\textit{San José State University}\\
San José, USA \\
lawrence.nguyen@sjsu.edu}
\and
\IEEEauthorblockN{Hiu Yung Wong}
\IEEEauthorblockA{\textit{Department of Electrical Engineering} \\
\textit{San José State University}\\
San José, USA \\
hiuyung.wong@sjsu.edu}
}

\maketitle

\begin{abstract}
This work presents a study of an implementation of a novel Quantum Convolutional Neural Network (QCNN) for binary classification of images from the Modified National Institute of Standards and Technology (MNIST) dataset. Using a novel architecture inspired by previous QCNN and classical convolutional neural network (CNN) implementations, we use a hierarchical partitioning approach to implement a QCNN circuit that can be approximated and simulated efficiently on a classical machine for a large problem. First, the original image is partitioned such that each process handles a smaller portion of the image, which is encoded into independent states. Then, these partitions merge and combine, resulting in states that contain information from both partitions while halving the number of processes. After repeating this until one process remains, we reduce the dimensionality of the state until a single qubit remains for measurement. Using this approach, we can use multiple processes in parallel to simulate a large QCNN program without the need for exponentially growing hardware requirements as the number of qubits increases. In our work, we use this scheme to train a 128-qubit model, which is impossible to run on any classical supercomputer without the novel architecture. We also explore the impact of this new model architecture on prediction accuracy by training it to perform binary classification on the MNIST dataset with a small number of qubits, and comparing it to a model without partitioning. Our initial findings show that partitioning images into smaller sub-images with this architecture does not degrade the model's performance and sometimes even improves it, likely because it reduces the Barren plateaus issue in the partitioning process.
\end{abstract}


\section{Introduction}

\subsection{Motivation}
As the field of quantum computing continues to mature, quantum neural networks are becoming an increasingly attractive research topic. QCNNs are an example of a quantum neural network that takes many design inspirations from classical CNNs, and they use parameterized quantum circuits to classify images \cite{b11}. In a previous study, researchers trained QCNN models to classify handwritten digits with using a system of 49 qubits, achieving an accuracy of 96\% using quantum hardware \cite{b9}. Simulating this number of qubits on a classical machine requires large amounts of memory, as seen in another study that simulated a 61-qubit quantum circuit using 768 terabytes of memory \cite{b14}, and it is almost impossible to scale to 128 qubits in the near future. When limited classical hardware is available, simulating QCNNs can serve only as educational examples with little practical value. For example, the QCNN example provided in the Qiskit machine learning tutorial uses an approach that encodes each pixel of the image being classified into a qubit \cite{b1}, and only 8 qubits are used to encode 8 pixels (2$\times$4 grid). However, if a QCNN is to ever be useful, it must be able to classify images larger than a 2$\times$4 grid of pixels, such as the 28$\times$28 handwritten digits in the MNIST dataset \cite{b13}. Encoding each pixel into a single qubit becomes an unfeasible strategy unless we drastically reduce the size of the image to be classified, which carries the risk of decreasing the accuracy of the model due to the loss of pixel information. 

Previous studies involving both QCNNs and CNNs have employed parallelization schemes to partition larger problems into smaller subproblems \cite{b15}\cite{b16}. In the QCNN study, it continuously splits the circuit into smaller sub-branches of circuits, reducing the dimensionality of the problem and running all circuits in parallel. Therefore, for a 128-qubit problem, it still needs to solve a 128-qubit quantum circuit at the beginning. In the study involving the classical CNN model, the architecture partitions the original image into smaller sub-images, with each partition training a "local" CNN. The outputs of these CNNs are then used as inputs for a dense neural network (DNN) which is trained to classify the original image. Taking inspiration from such studies, we attempt to alleviate these constraints by parallelizing the circuit across different processes in a High-Performance Computing (HPC) cluster. We use the same partitioning strategy as the CNN-DNN architecture \cite{b15}, but we opt to combine our processes via a binary reduction tree after each convolutional block rather than training a DNN at the end. Towards this we introduce a novel architecture that allowed us to simulate 128 qubits on minimal classical hardware without degrading the overall performance of the model.

\subsection{Paper Contents and Organization}
In section II, we discuss the software libraries and physical hardware we used in this work. Then, we describe the MNIST dataset. We also briefly discuss the the basic architecture of a CNN and the components of the Qiskit tutorial that we are building upon. In section III, we discuss the implementation of our parallel structure. First, we describe how we partition the image to be classified to different processes. Then, we describe how these partitions manipulate the data and eventually interact with the other partitions. Next, we discuss the changes we implemented to how our parameters are defined to support this new structure. Finally, we touch on how we implemented the training loop use to fine-tune all of our new parameters. Section IV describes the results of our experiments using this structure, and Section V is a brief conclusion and speculates on future options for continuing our work. 

\section{Background and Setup}
\label{sec: II. Design and Device Specifications}

\subsection{Tools and Software}
All of the code produced was run on a High-Performance Computing (HPC) cluster. The cluster has a partition in which we are able to utilize nodes equipped with NVIDIA H100 96GB HBM3 GPUs, AMD EPYC 9454P CPUs, and up to 512GB of memory. In this paper, we only used the CPUs to demonstrate the power of the novel architecture in minimizing resource requirements. For our 64-qubit simulation, we split the work across 16 processes: 4 nodes, 4 processes per node. Similarly for our 128-qubit simulation, we split the work across 32 processes: 4 nodes, 8 processes per node. Each process is assigned 8 gigabytes of memory. We construct all of our circuits and simulate the quantum state evolution using Qiskit (version 1.4.4), and we use MPI4PY (version 4.1.1) for parallelization.

\subsection{Dataset}
\begin{figure}[H]
    \centering
    \includegraphics[width=\linewidth]{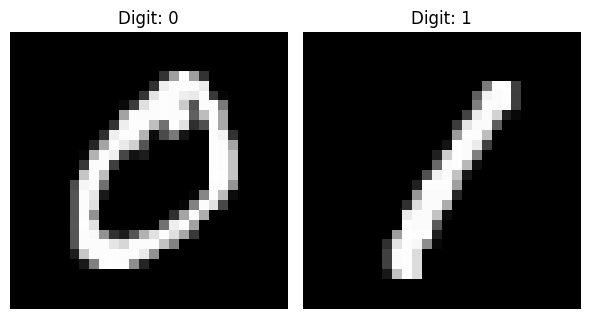}
    \caption{Data samples from MNIST}
    \label{fig:mnistexample}
\end{figure}

The MNIST dataset is a widely used dataset used to train and evaluate image related machine learning tasks. It is composed of images of a single hand written digit that is displayed in a 28$\times$28 pixel image. The digits in this dataset range from 0 to 9, all written with varying handwriting styles. In total the dataset has 60,000 training images and 10,000 test images. For the purposes of this study we have opted to only perform a binary classification task, classifying an image between one of two possible digits: 0 and 1. For this study, we use a small subset of this dataset with 125 samples total: 63 0s and 62 1s with a 80:20 train test split. Fig.~\ref{fig:mnistexample} displays sample images of these digits in their original resolution. 

\subsection{Classical CNN Background}
A CNN is a classical neural network that is primarily used to perform classification tasks on images \cite{b12}. They are primarily composed of two types of parameterized layers: convolutional and pooling. The convolution layer’s primary function is to perform feature extraction by applying a sliding kernel across a smaller portion of the original image. These kernels produce feature maps, which represent where in the image certain features were found. The pooling layer reduces the dimensionality of these feature maps, condensing the information into a smaller feature space. These layers are implemented together in an alternating manner to reduce the original image into a vector that the algorithm will predict on. These parameterized layers are fed through an optimization loop, allowing the model to adjust its parameters and eventually learn relevant information pertaining to its classification task.

\subsection{QCNN Background}
The QCNN architecture draws inspiration from its classical counterparts but implements these ideas using quantum circuits \cite{b1}. The one upon which our novel architecture is built is based on a Qiskit tutorial. These circuits are parameterized 2-qubit unitaries that emulate the behavior of their classical counterparts by learning the relationship between paired qubits through shared information. The inputs of these circuits are individual pixels from the image to be classified that have been encoded into a quantum state via Qiskit's Z-feature map \cite{b1}. 

\begin{figure}[H]
    \centering
    \includegraphics[width=.5\linewidth]{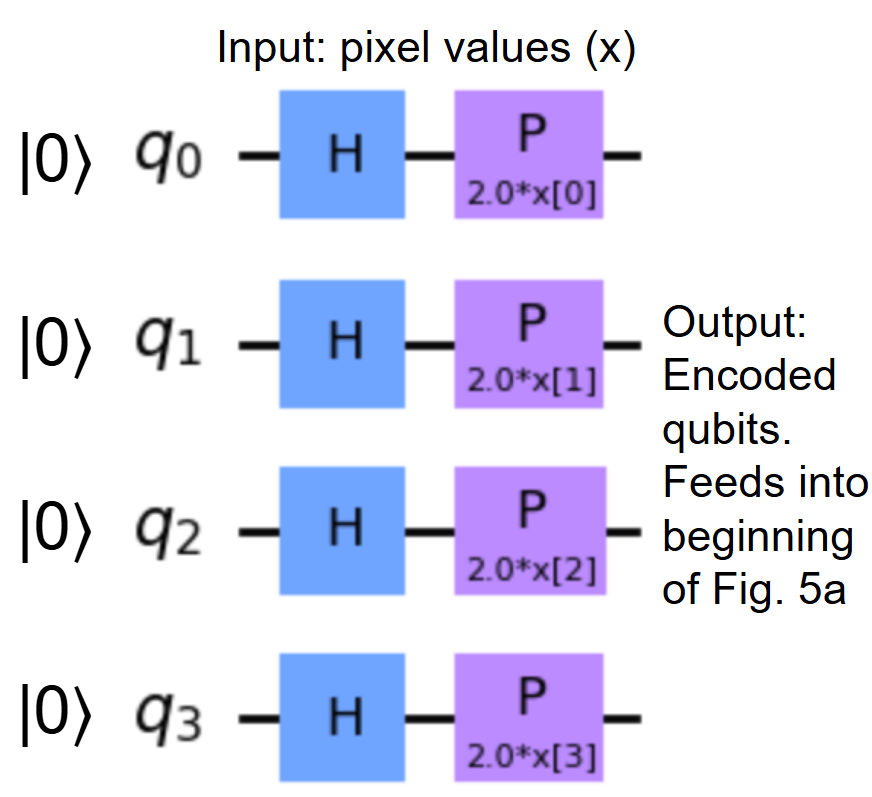}
    \caption{Qiskit's Z-FeatureMap on 4 qubits. They are all initialized to $\ket{0}$ followed by a parameterized phase gate, $\bm{P(x)}$.The angle being rotated is based on the pixel value, $x$. The output is fed into the first convolution layer of Fig.~\ref{fig:full_overview}a.}
    \label{fig:featuremap}
\end{figure}

\begin{figure}[H]
    \centering
    \includegraphics[width=\linewidth]{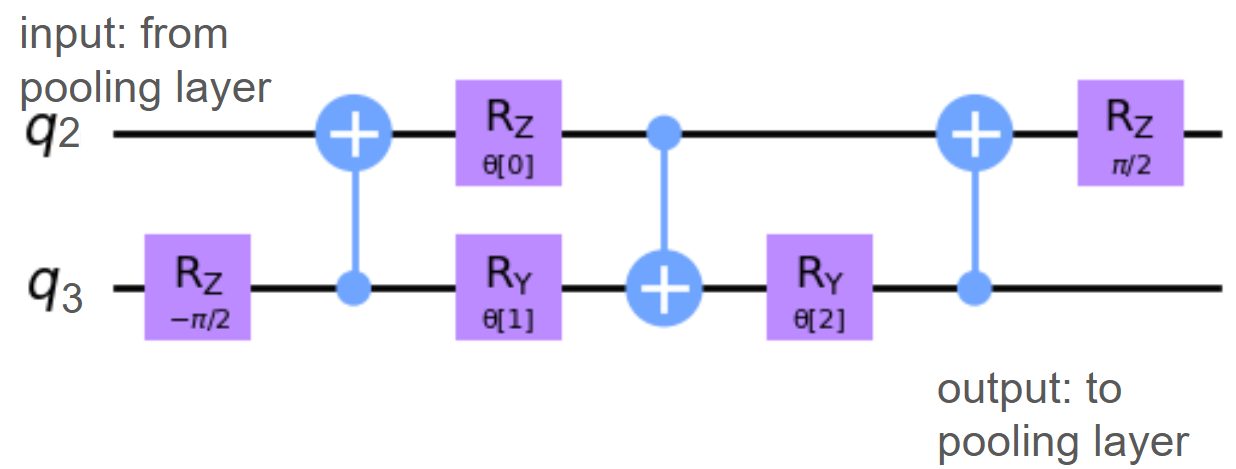}
    \caption{Convolutional circuit that serves as the building block for a convolutional layer (Fig.~\ref{fig:full_overview}). This is the second convolutional layer in Fig.~\ref{fig:full_overview}a. The circuit mixes information and prepares the states to be pooled by pooling layers. \cite{b1}}
    \label{fig:convolutionlayer}
\end{figure}

The Z-feature map is a 1-qubit quantum circuit found within Qiskit's circuit library that encodes a real number, denoted as $x$, into a quantum state (Fig.~\ref{fig:featuremap}). The 2D array of pixels is flattened into a single dimension, and each pixel is assigned to a qubit which has been initialized to $\ket{0}$. The Z-feature maps apply the Hadamard gate to each qubit, and then use a phase gate, denoted as $\bm{P(x)}$ in Fig.~\ref{fig:featuremap}, to rotate the state proportional to the pixel value, $x$, so as to encode the pixel value into a quantum state, which will be fed into the following convolution layer.
The convolutional circuit, pictured in Fig.~\ref{fig:convolutionlayer}, is modeled after the most general form of a 2-qubit unitary \cite{b2}: 
\begin{equation}
    N(\alpha,\beta,\gamma) = exp(i[\alpha \sigma_x\sigma_x + \beta\sigma_y\sigma_y + \gamma\sigma_z\sigma_z])
\end{equation}
Each convolutional circuit has 3 trainable parameters, denoted by $\theta$ in Fig.~\ref{fig:convolutionlayer}, which are adjusted by a classical optimizer during the training loop. These parameterized gates are distributed across a 3-CNOT structure, which serves as a SWAP gate if there were no rotation gates in between. The presence of these parameterized gates controls how much information is actually mixed between the paired qubits. 

\begin{figure}[H]
    \centering
    \includegraphics[width=\linewidth]{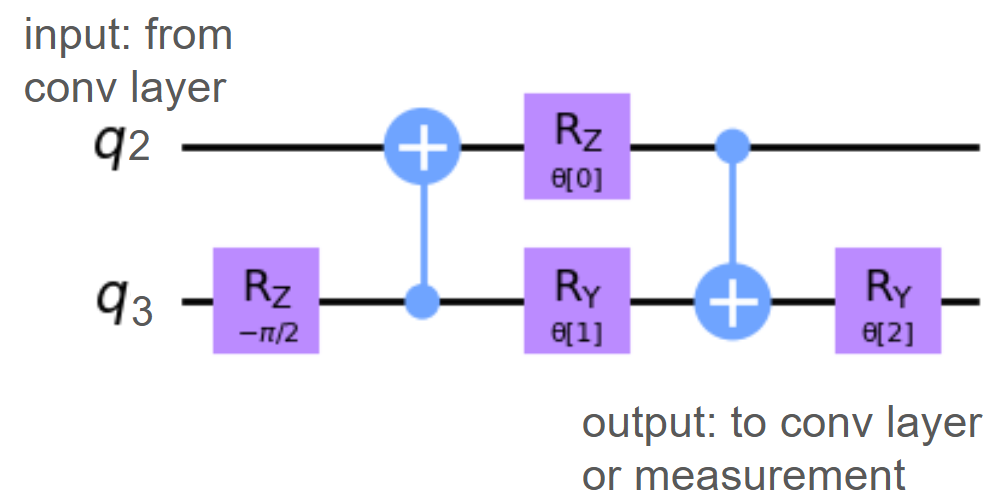}
    \caption{Pooling circuit and the last pooling layer of Fig.~\ref{fig:full_overview}. $q_2$ is the source qubit and $q_3$ is the sink qubit \cite{b1}.}
    \label{fig:Pooling}
\end{figure}

Similarly, a pooling circuit, pictured in Fig.~\ref{fig:Pooling}, is also a 2-qubit unitary that has 3 trainable parameters. These parameters, also denoted as $\theta$ (but are different parameters than those in the convolution layers), are also adjusted during the training loop. The main distinction between the two circuits is that the pooling circuit does not include the third controlled-NOT gate after the third parameterized gate execution. By doing this, the circuit is pooling the information from the “source qubit”, denoted by $q_2$ in the figure, into the “sink qubit”, which is denoted as $q_3$, such that all the relevant information resides in the sink qubit \cite{b1}. This is because of the lack of entanglement symmetry after the parameterized gates are applied. Since the parameterized gates in the sink qubit do not influence the source, all of the relevant information that the model identifies is condensed in the sink. After all of the information is condensed, the source can be ignored for the remainder of the circuit.

\begin{figure}
    \centering
    
\begin{minipage}[b]{\linewidth}
    \centering
    \includegraphics[width=\linewidth]{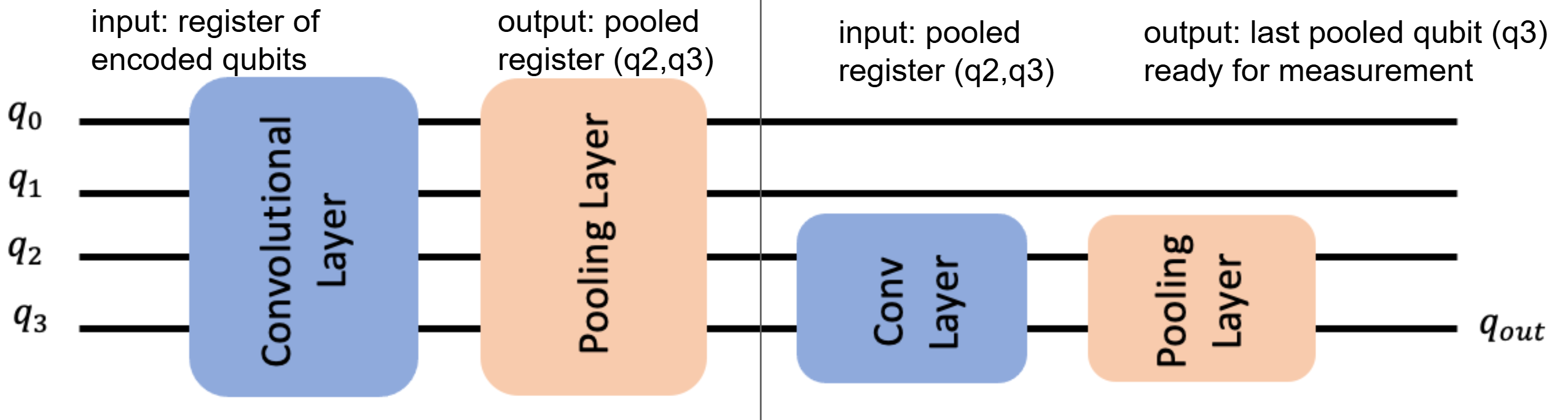}
    \small (a) An example of a 4-qubit QCNN Architecture \cite{b1}
    \label{fig:placeholder}
\end{minipage}

\vspace{1em}

\begin{minipage}[b]{\linewidth}
    \centering
    \includegraphics[width=\linewidth]{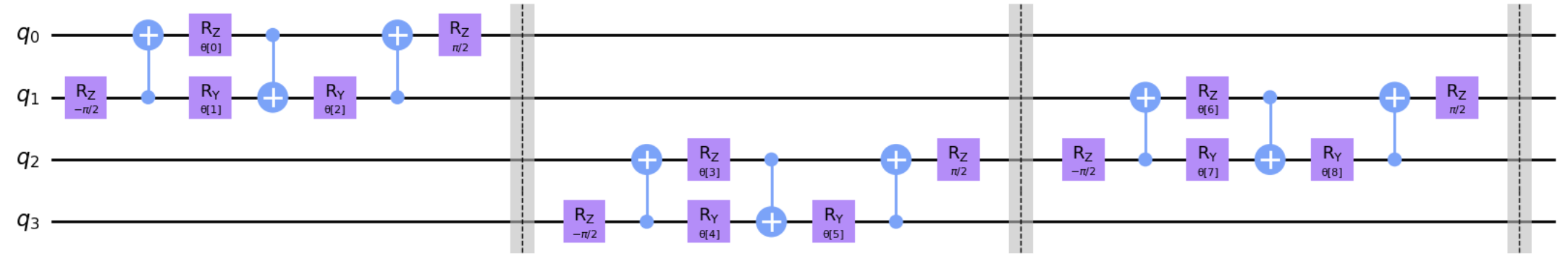}
    \small (b) 4-qubit convolutional layer, pairing all neighboring qubits: ($q_0$, $q_1$), ($q_2$, $q_3$), and ($q_1$, $q_2$). Used as the first convolutional layer in Fig.~\ref{fig:full_overview}a.
\end{minipage}

\vspace{1em}

\begin{minipage}[b]{\linewidth}
    \centering
    \includegraphics[width=\linewidth]{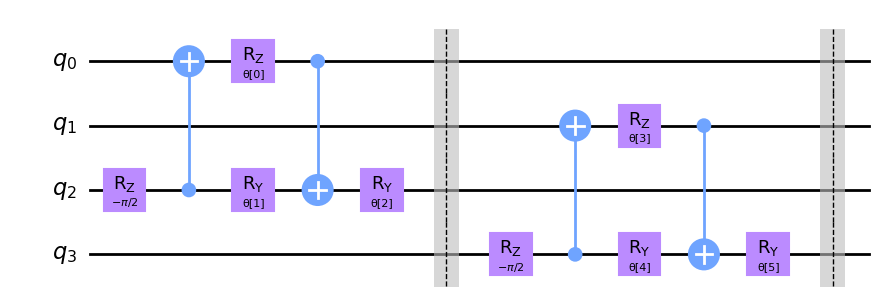}
    \small (c) 4-qubit pooling layer, $q_0$ and $q_1$ are the "source" qubits. $q_2$ and $q_3$ are the "sink" qubits. Used as the first pooling layer in Fig.~\ref{fig:full_overview}a.
\end{minipage}

 \caption{The architecture of a 4-qubit QCNN after 4 pixels have been encoded into quantum states via a Z-FeatureMap. The circuits of the last two layers are shown in Figs. \ref{fig:convolutionlayer} and \ref{fig:Pooling}, respectively. a) The overall structure of the architecture, highlighting how layers are alternated to reduce dimensionality. b) How the first convolutional layer is built by pairing up neighboring qubits and applying convolutional circuits. c) How the first pooling layer is built by designating sources ($q_0$ and $q_1$) and sinks ($q_2$ and $q_3$) for pooling circuits.}
     \label{fig:full_overview}
\end{figure}

Both of the aforementioned circuits are used as building blocks to create imitations of classical convolutional and pooling layers. Fig.~\ref{fig:full_overview} shows a 4-qubit QCNN after Z-FeatureMap. To form the convolutional layers similar to the classical CNN counterparts, the convolutional circuits are applied on neighboring qubits such that each qubit is paired with the qubit above and below it on the register. The Qiskit tutorial also pairs the first and last qubits together, but we do not in our architecture. Similarly, pooling layers are formed by pairing off the qubits into a register such that half of the qubits are designated as sources and the other half as sinks, as seen in Fig.~\ref{fig:full_overview}c. If there is an odd number of qubits, two qubits act as the source for a single sink qubit, executing the circuit twice with different sources, but the same sink. Doing so means that after the execution of every pooling layer, the architecture cuts the number of relevant qubits in the system in half, rounding down if necessary. Applying these layers in an alternating manner, as seen in Fig.~\ref{fig:full_overview}a, extracts features and reduces the system until only a single qubit remains. The Pauli Z expectation value of this last qubit is used in the model's prediction. The possible measurement outcomes are -1 or 1. The expectation is thus a value ranging from -1 to 1, which is mapped to one of two labels in a binary classification problem. For example, a positive expectation value corresponds to digit 1, and other values correspond to digit 0. 

\section{Novel Architecture}
\label{sec: III. Methodology}

Given a large number of pixels, the problem becomes being able to represent all the information within a qubit register. A large number of qubits is unrealistic to simulate with a single process, as the size of the matrices scales exponentially and becomes unmanageable for the physical memory of the machine. It is known that even with supercomputers, it is not feasible to simulate a system with more than 50 qubits without approximations. To remedy this problem, we have devised a novel parallel architecture that can be implemented in real hardware for QCNN and, at the same time, can be simulated with classical computers.

\subsection{Partitioning}
\begin{figure}[]
    \centering
    \includegraphics[width=\linewidth]{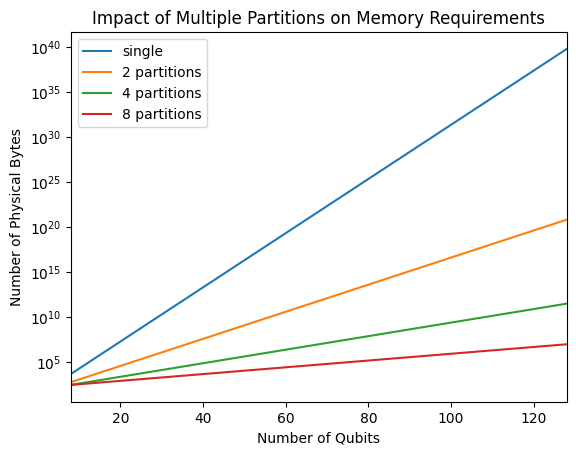}
    \caption{Physical memory requirement to store a state as a function of the number of qubits and partitions}
    \label{fig:scaling_plot}
\end{figure}

Before building the parallel circuit structure, our first step is to partition the original image into smaller independent sub-images. By dispersing the original $N$-qubit problem into $M$ partitions, we can assign each partition $N/M$ qubits (assume divisible). Each partition is then given a separate process to run on, each with its own allocated amount of memory, and treated as independent of the others. Without partitioning, the number of physical bytes, denoted as $P_{bytes}$, needed to represent $N$ qubits can be estimated as the following \cite{b3}:
\begin{equation}
    P_{bytes} = 2^{N+4} 
\end{equation}
This is because each state vector in Qiskit is represented as an array of $2^N$ complex numbers which use the complex128 data type, or 16-byte complex numbers \cite{b17}. However, using the partitioning scheme, our physical memory requirements become: 
\begin{equation}
    P_{bytes} = (2^{(N/M)+4})M
\end{equation}
The impact of partitioning on physical memory requirements can be seen in Fig.~\ref{fig:scaling_plot}. We can see that partitioning qubits into 2 processes drastically impacts the amount of memory required.

\begin{figure}
    \centering
    
\begin{minipage}[b]{\linewidth}
    \centering
    \includegraphics[width=\linewidth]{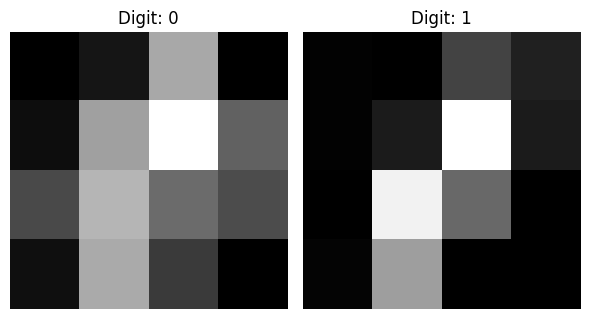}
    \small (a) Resized images from Fig.~\ref{fig:mnistexample} for a 16-qubit configuration.
\end{minipage}

\vspace{1em}

\begin{minipage}[b]{\linewidth}
    \centering
    \includegraphics[width=\linewidth]{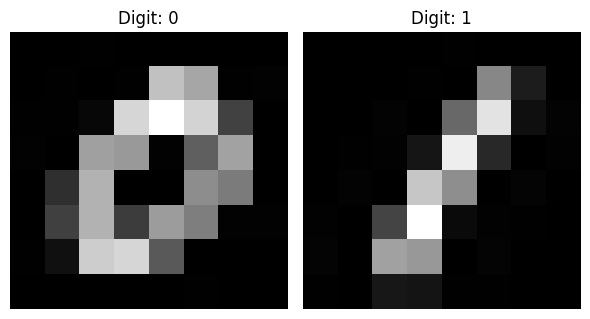}
    \small (b) Resized images from Fig.~\ref{fig:mnistexample} for a 64-qubit configuration.
\end{minipage}

\vspace{1em}

\begin{minipage}[b]{\linewidth}
    \centering
    \includegraphics[width=\linewidth]{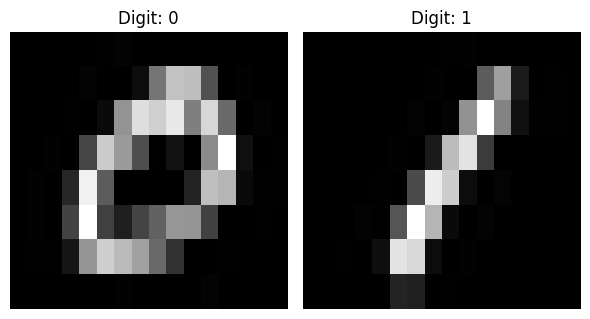}
    
    \small (c) Resized images from Fig.~\ref{fig:mnistexample} for a 128-qubit configuration.
\end{minipage}

 \caption{Resized images based on the total number of the final pixels (qubit used). a) 16 pixels organized into a 4$\times$4 image. b) 64 pixels arranged into an 8$\times$8 image. c) 128 pixels organized into an 8$\times$16 image.}
 \label{fig:downsized_digits}
\end{figure}

\begin{figure}
    \centering
    \includegraphics[width=\linewidth]{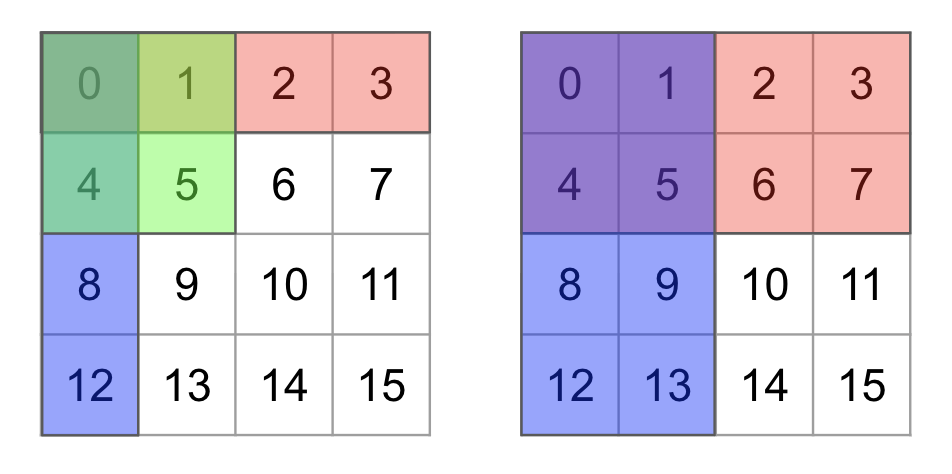}
    \caption{Different grouping methods for partitioning configurations. Each box represents a single pixel, with the number indicating pixel index. \textbf{Left:} 4 4-qubit partitions. Green is split into quadrants, with 2$\times$2 pixels each. Blue is split by columns and red is split by rows. \textbf{Right:}2 8-qubit partitions. Blue is split by columns and red is split by rows.}
    \label{fig:groups1}
\end{figure}

\begin{figure}
    \centering
    \includegraphics[width=\linewidth]{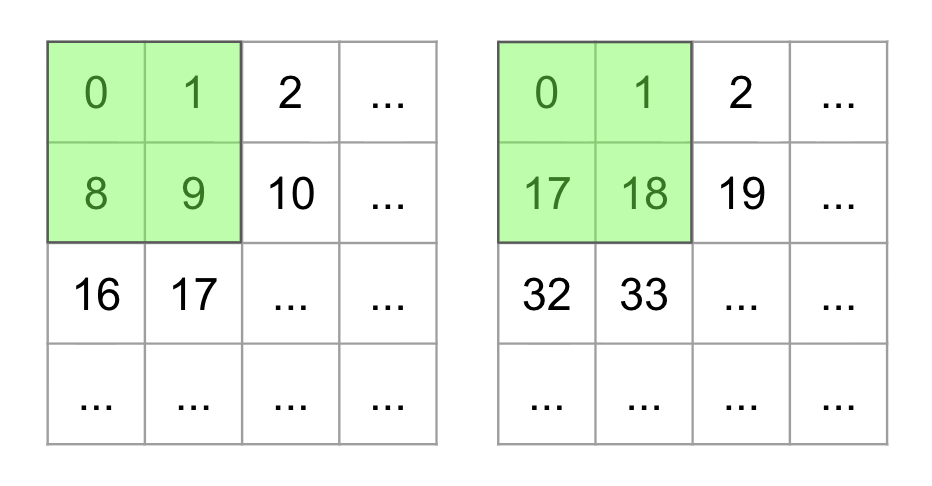}
    \caption{The grouping method for a large number (16 and 32) of 4-qubit partitions. Both configurations group pixels into 2$\times$2 squares. Each box represents a single pixel, with the number indicating pixel index. \textbf{Left:} 16 4-qubit partitions for a 64-qubit problem. \textbf{Right:} 32 4-qubit partitions for a 128-qubit problem.}
    \label{fig:groups2}
\end{figure}

To partition our image into smaller sub-images, we first use a classical average pooling downsizing method to resize the original image to have the same number of pixels as there are total qubits, $N$. Fig.~\ref{fig:downsized_digits} displays downsized versions of the digits displayed in Fig.~\ref{fig:mnistexample} in a 16-qubit, 64-qubit system, and 128-qubit system. The pixel dimensions of the downsized image preserve the dimensions of the original image as closely as possible. For example, in Fig.~\ref{fig:downsized_digits}c, we downsize the image into an 8$\times$16 image because 128 pixels cannot fit into a perfect square. These downsized images are then grouped into smaller chunks and distributed to different processes. For smaller systems, such as a 16-qubit configuration, we can group these pixels in different configurations: rows, columns, and quadrants. Figs.~\ref{fig:groups1} and~\ref{fig:groups2} display the possible ways to distribute the downsized images into sub-images, where each box represents a single pixel, and the number in the box is the pixel's index. We try different grouping methods to see if the spatial context that each grouping method provides has an impact on the resulting accuracy of the model. In the case of a 64-qubit system, we use a configuration of 16 4-qubit processes with each process handling a 2$\times$2 pixel chunk. In the case of a 128-qubit system, we use a similar configuration of 32 4-qubit processes, with each process handling a 2$\times$2 pixel chunk. After each process is assigned a subset of pixels from the downsized image, each group of pixel values is flattened and we encode these values into the qubits with a Z-feature map.

\begin{figure*}[t]
    \centering
    \includegraphics[width=\linewidth]{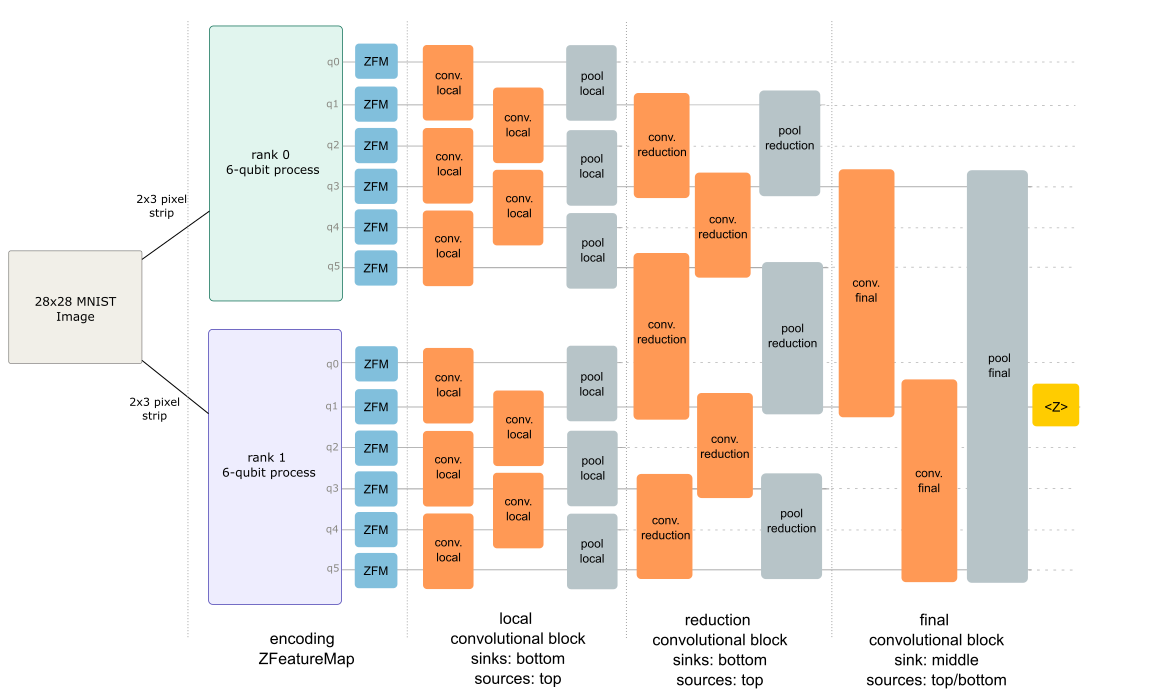}
    \caption{The novel architecture of the QCNN with 12 qubits with 2 partitions. Convolutional and pooling layers are alternatively applied to the circuit. This condenses the information and reduces the system to a single qubit, which is measured for a classification.  Qubits that are ignored are represented as dotted lines in this diagram.}
    \label{fig:arch_mock}
\end{figure*}
\subsection{Reduction and Combination Architecture}

The circuit that can be run on a real quantum computer is shown in Fig.~\ref{fig:arch_mock}. It shows an example of partitioning a 12-qubit problem into two groups of 6-qubit problems. After partitioning and encoding the qubits, we apply a distributed QCNN architecture to train the model. First, each independent system applies a single convolutional and pooling layer to the their qubits. Similar to previous implementations and classical counterparts, these layers condense the relevant information into half of the partition's original amount of qubits. After each combination stage, half of the qubits are not used (equivalent to applying identity gates). However, to simulate this circuit, the resources will grow exponentially with the number of qubits.

\subsection{Simulation Strategy}

\begin{figure*}[t]
    \centering
    \includegraphics[width=\linewidth]{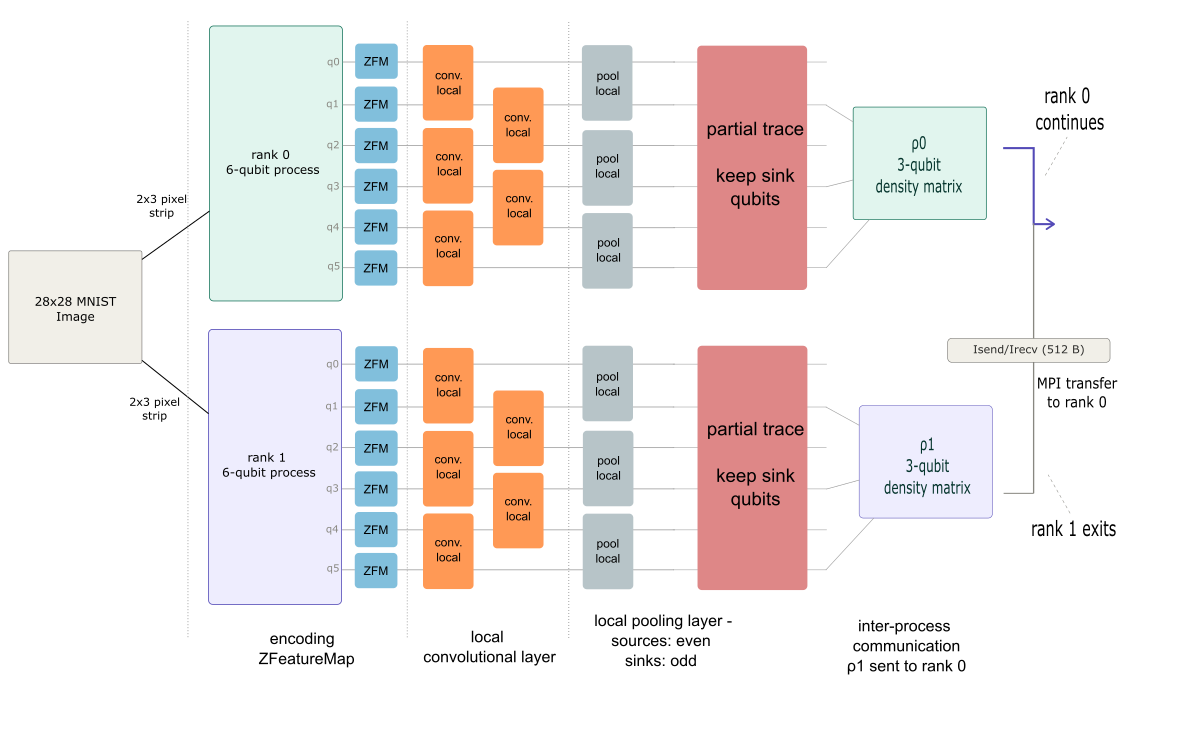}
    \caption{The first half of the simulation circuit for the physical circuit in Fig.~\ref{fig:arch_mock}. This is for a model being trained with 2 6-qubit processes. Images are partitioned then encoded. After the first convolutional and pooling layers, the source (evens) qubits are traced out, leaving 3-qubit density matrices. All parameterized circuits use local parameters. Processes are then paired up such that one of the processes holds both density matrices, preparing it for the combination step.}
    \label{fig:arch1}
\end{figure*}

Since we no longer care about the source qubits in each of the partitions, we can represent each partition in a reduced state by taking the partial trace over the source qubits in simulations. This is achieved by converting the state vectors into density matrices and then tracing out the source qubits, which we designate as the even-numbered qubits in our architecture, leaving us with only the density matrix of the sink qubits \cite{b4}.
\begin{equation}
    \rho_{sink} = Tr_{source}(\rho_{total})
\end{equation}
Then, neighboring processes are paired with each other by process order. One of the paired processes receives its partner's density matrix while the other sends it, as seen in the inter-process communication step in Fig.~\ref{fig:arch1}. The new density matrix for a new state is then created via the result of a tensor product of both processes' reduced density matrices as denoted below.
\begin{equation}
    \rho_{combined} = \rho_{send} \otimes \rho_{recieve}
\end{equation}
This is possible because both states, prior to combination, were completely independent from each other. A process that shared its density matrix with its partner is then stopped, as all of its information is contained in the new state carried by its partner. \emph{It should be noted that, after the first combination, we are simulating the density matrix, \bm{$\rho$}, instead of state vectors, $\ket{v}$. Therefore, the evolution of the density matrix is used. Instead of calculating $\bm{U}\ket{v}$, $\bm{U}\bm{\rho}\bm{U}^\dagger$  is calculated with the evolve() function in Qiskit.} The resulting remaining states are still $N/M$ qubits in size, however the number of total processes is halved. We then apply a convolutional and pooling layer to reduce the system to prepare it for the next reduction, repeating these steps, or final classification steps.

\begin{figure}
    \centering
    \includegraphics[width=\linewidth]{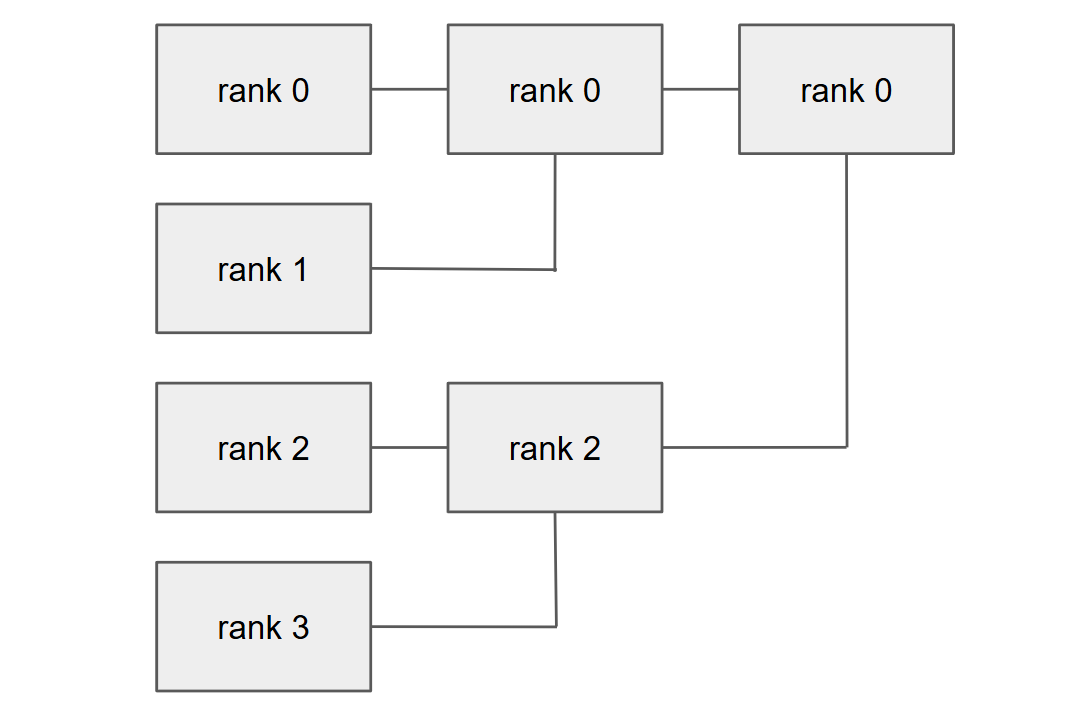}
    \caption{A 4-process example binary tree reduction showcasing how processes are paired, combined. When the last process (rank 0) is the only process remaining, we apply our final convolutional and pooling layers.}
    \label{fig:tree_reduciton}
\end{figure}

This reduction pattern, which is displayed in Fig.~\ref{fig:tree_reduciton} for a model being trained with 4 processes, is repeated until there is only a single process remaining, mimicking a binary reduction tree structure. This introduces the constraint that the number of defined processes must be a power of 2. Then, exactly like our initial model, we apply alternating convolutional and pooling layers on the remaining state until a single qubit remains for measurement and classification. This process emulates the dimensionality reduction seen in classical CNNs. 

\begin{figure*}[]
    \centering
    \includegraphics[width=\linewidth]{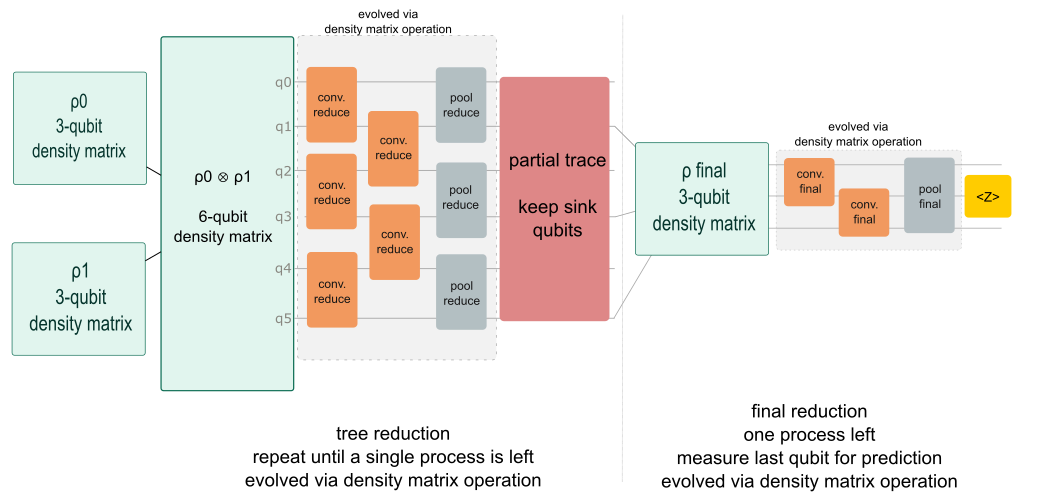}
    \caption{The second half of the simulation circuit for the physical circuit in Fig.~\ref{fig:arch_mock}. After the density matrices in parallel processes are prepared and shared, they are combined via a tensor product. After this combination, we apply a convolutional and pooling layer to this combination to reduce the state into another 3-qubit density matrix. These gates use the reduction parameters. Then this density matrix is paired with another from another process and we repeat this cycle, using a binary tree reduction. When one process remains, we reduce the system to a single qubit using the final parameters, then measure for our prediction. All gates in this diagram are applied via the DensityMatrix evolve() method from Qiskit.}
    \label{fig:arch2}
\end{figure*}

Therefore, for the 12-qubit example in Fig.~\ref{fig:arch_mock}, the simulation structure is depicted in Fig.~\ref{fig:arch1} and~\ref{fig:arch2} with details. 


\subsection{Parameterization}
The parallelized nature of the new architecture changes the way parameters are used within the system. Our structure contains three distinct groups of parameters: local, reduction, and final. Here, local refers to the first stage of the QCNN. Final refers to the last stage of the QCNN. Reduction refers to all stages in between where dimensionality is reduced. In our parallel structure, every independent process has an identical set of local parameters. By applying the same local parameters on each partition of the original image, we are effectively emulating the behavior of a classical CNN which applies the same sliding kernel across the image to extract features. These parameters can be seen in the local convolutional layer in Fig.~\ref{fig:arch1}. Reduction parameters, seen in the tree reduction section in Fig.~\ref{fig:arch2}, are used as the convolutional and pooling circuit parameters after a combination step has occurred. These parameters are distinct across each tree reduction layer, because their goal is to extract information between combinations of states at various levels of combination. The final parameters, seen in the final reduction portion of Fig.~\ref{fig:arch2}, are all of the parameters that remain after the final combination and reduction steps are completed. These parameters are only used by the last process in our system. All of these parameters are randomly initialized in a uniform distribution centered around 0 and range between -0.1 and 0.1.

Due to the aforementioned parameter definitions of the quantum circuit, the total number of parameters is dependent on both the number of qubits in each process and the number of processes we have defined. Each individual convolutional and pooling circuit has 3 parameters. A convolutional layer creates $n-1$ pairs of qubits, where $n$ is the number of qubits in the layer. When it is divisible, $n=N/M$. The number of parameters of each $n$-qubit convolutional layer is $3(n-1)$. A pooling layer creates $n/2$ pairs, meaning that the number of respective parameters is $3(n/2)$. This means that the number of parameters, denoted as $P$, in a convolutional block, or the combination of a convolutional and pooling layer, of $n$ qubits can be estimated as the following.
\begin{equation}
    P(n) = 3(n - 1) + 3(n/2) = (9n/2)-3
\end{equation}
Therefore, the number local parameters, for a process with $n$ qubits can be written as
\begin{equation}
    P_{local} = P(n)
\end{equation}
Because the local parameters are shared across all processes and the final parameters are only used on the final process remaining, the only impact the number of processes has on the number of parameters is how many sets of reduction parameters are stored, i.e.,  $log_2(M)$. The reduction circuit is architecturally the same as the local circuit, so the total number of reduction parameters, given $M$ number of processes, can be written as the following.
\begin{equation}
    P_{reduction} = log_2(M)P(n)
\end{equation}
The number of final parameters is just the summation of parameters for each convolutional block needed to reduce the last process down to a single qubit. Since the reduction layer returns a state with $n/2$ qubits, we can express this as a summation of number of parameters in convolutional blocks, P, for every power of 2 until $n/2$:

\begin{equation}
    P_{final} = \sum_{k=1}^{log_2(n)-1}P(2^k)
\end{equation}
In the special case of the number of qubits not being a power of two, such as the 3 remaining qubits in Fig.~\ref{fig:arch2} we just need to consider that there is an extra pooling circuit execution to pool 3 qubits into a single one. The total number of parameters in a system $M$ processes, each with $N$ qubits can be is just the sum of all the types of parameters:
\begin{equation}
    P_{total} = P_{local} + P_{reduction}+P_{final}
\end{equation}
Table 1 shows how the number of parameters scales within our architecture.

\begin{table}[H]
\centering
\renewcommand{\arraystretch}{1.3} 
\setlength{\tabcolsep}{10pt}      

\begin{tabular}{lccc}
\toprule
\textbf{Qubits per process} & \textbf{1 process} & \textbf{2 processes} & \textbf{4 processes} \\
\midrule
4 qubits  & 21  & 36  & 51  \\
8 qubits  & 54  & 87  & 120 \\
16 qubits & 123 & 192 & 261 \\
\bottomrule
\end{tabular}

\vspace{1em}
\caption{Parameter counts across varying qubit and process counts}
\label{tab:qcnn_params}
\end{table}

\subsection{Training Loop}
The training loop we have implemented is a stochastic gradient descent. This is a classical parameter optimization algorithm that works by updating a model's parameters based on a gradient calculated after a single pass of a sample \cite{b6}. The loss function we have chosen is a simple mean-square error, and we calculate our gradient by using the parameter shift rule \cite{b7}. The parameter shift rule allows us to calculate the derivative of our expectation value, or predicted value $f(\theta)$, by performing the same circuit twice with the parameter shifted up and down:
\begin{equation}
    \frac{\partial f}{\partial \theta}= \frac{f(\theta + \pi/2)-f(\theta -\pi/2)}{2}
\end{equation}
However, this means that a forward pass must be conducted twice for a single parameter to be updated, which doubles the amount of training time. In order to update all of the parameters after a single sample, we would need to perform several forward passes to update all of them, which is extremely time consuming. To save time, we have opted to adjust 15 parameters per sample in the training set only. Therefore, 1500 parameters are adjusted per epoch if there are 100 training samples. To do so, we created a parameter index that contains all of the parameters initialized for the model. At the start of each epoch, we randomize the order of this index. The parameters updated per sample are chosen in this order, looping back around after all parameters have been updated. 

\subsection{Adjustments and Considerations}
During the development and testing of this architecture, we found that it takes systems with larger qubit counts per process a significantly longer time to train a model, i.e. 4 16-qubit processes takes significantly longer than 16 4 qubit processes. This is likely due to the exponential growth of computational power needed to perform the matrix operations needed to simulate the states and density matrices as the system does its forward passes. This, compounded with the fact that several forward passes are made each sample makes up the drastic differences in runtime. This is why we have elected to only pursue 4-qubit processes, grouped in 2$\times$2 pixel chunks, when attempting to simulate our larger configurations of 64 and 128 qubits in the following Section.


\section{Results}
\begin{figure}
    \centering
    \includegraphics[width=\linewidth]{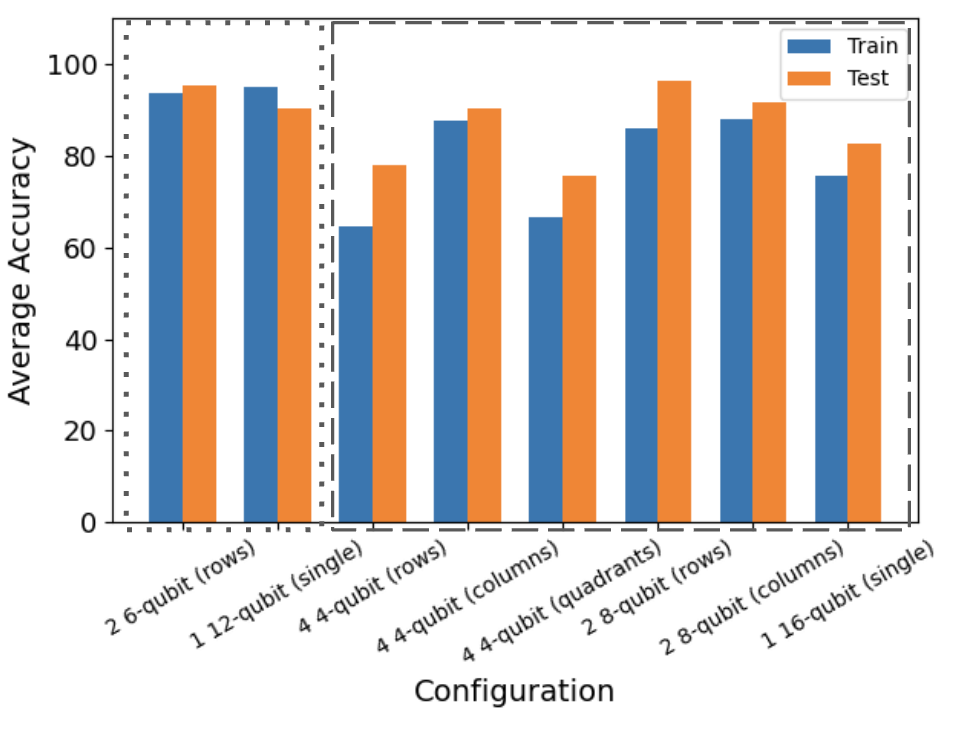}
    \caption{Average accuracies of single vs parallel configurations. The accuracies boxed in the dotted lines are the results for the 12-qubit problem and the accuracies boxed in the dashed lines are for the 16-qubit problem.}
    \label{fig:avg1}
\end{figure}

To test the effectiveness of this new architecture, for the same problem, we trained several models and compared their accuracies. First, we trained 10 models using a single 12-qubit process and another 10 using a 16-qubit process. These models represent typical QCNNs without parallelization. Then, we trained 10 additional models for each possible grouping with our parallel architecture. We trained multiple models to average out the noise introduced by the order of parameter updates in the training loop and random parameter initialization. All of these models were trained using 100 samples over 100 epochs with a learning rate of 0.1. All models were tested with an additional 25 samples sourced from the train test split we discussed earlier. These low sample counts were used in the name of expediency, as we have not implemented any graphics card acceleration for this architecture, and because the models trained display sufficient accuracies. While other studies have found 50 epochs to be sufficient to train an effective model, we elected to train over 100 epochs to allow our models more time to converge \cite{b10}. The averages for each configuration can be seen in Fig.~\ref{fig:avg1}.

Comparing the average accuracies of configurations with the same number of qubits, we can see that the parallel architecture performs competitively. Models trained with 2 6-qubit processes perform as well as models trained with a single 12-qubit process. In other cases, we see some mixed results. Models trained with 2 8-qubit configurations and models trained with 4 4-qubit models grouped into columns seem actually outperform a their single 16-qubit counterpart in both training and test accuracy. However, models trained with 4 4-qubit processes grouped into rows or quadrants seem to perform worse than their single process counterpart. This implies that the way these partitions are grouped impacts the overall accuracy of the model. 

\begin{figure}
    \centering
    \includegraphics[width=\linewidth]{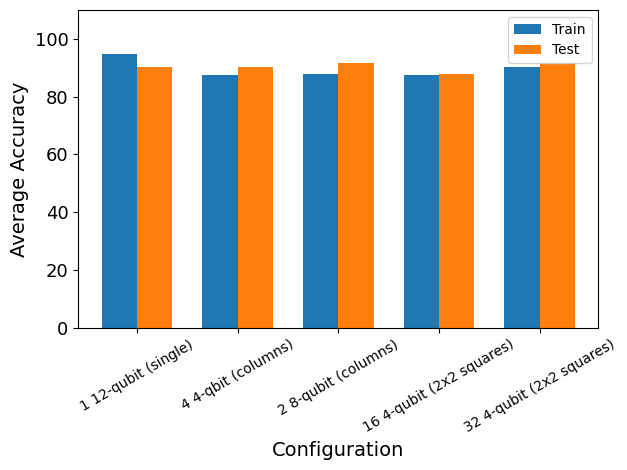}
    \caption{Average accuracies of the best performing models trained with low qubit counts compared to models trained with 64 and 128 qubit systems.}
    \label{fig:avg2}
\end{figure}

Next, we used same approach to train much larger systems, but partitioned into 4-qubit processes due to time constraints. We trained 10 models using 16 4-qubit processes, which is a total system of 64 qubits, and another model using 32 4-qubit processes, totaling to 128 qubits. The original image for the 128-qubit case is downsized to 8$\times$16 pixels. The average accuracies of these larger models are compared to the best performing smaller models: a single 12-qubit process, 4 4-qubit processes grouped into columns, and 2 8-qubit processes grouped into columns. This comparison can be seen in Fig.~\ref{fig:avg2}.
Here, we can clearly see that the models trained with a large number of qubits performs similarly to the best performing configurations trained on smaller systems. Despite being grouped by 2$\times$2 pixel squares, which is one of the worse grouping configurations seen in Fig.~\ref{fig:avg1}, larger models are able to achieve competitive training and testing accuracies when compared to their smaller counterparts. This may signify that adding more qubits to the system, and therefore more pixels to the downsized image, does increase the overall accuracy of the model.

\section{Conclusion}
This study has shown that it is possible to parallelize simulated QCNNs support several times more qubits than what is possible in a single process given the same amount of total memory. By using the novel architecture that we have introduced, we have shown that we can train competitive QCNN models using both 64-qubit and 128-qubit systems using classical computers. When pixels are grouped in certain configurations, this parallel approach produces models with higher the accuracy scores when compared to accuracies of models trained on a single process with the same number of total qubits. These results suggest that further testing, exploration and refinement of this architecture is well justified.

\section{Acknowledgment}
This work was supported in part by the National Science Foundation under Grants 2125906 and 2430291.


\end{document}